\documentclass[12pt]{article}

\usepackage{amsmath,amssymb}
\usepackage{authblk}
\title{ Quantal Response Equilibria in Binary Choice Games on Graphs}
\date{}
\author{Andrey Leonidov$^{(a,b)}$}
\author{Alexey Savvateev$^{(b,c)}$}
\author{Andrew G. Semenov$^{(a,d)}$}
\affil{{\small 
(a) P.N. Lebedev Physical Institute, Moscow, Russia\\
(b) Moscow Institute of Physics and Technology, Dolgoprudny, Russia\\
(c) Central Economics and Mathematics Institute, Moscow, Russia\\
(d) National Research University Higher School of Economics, Moscow, Russia
 }}

\begin{document}
\maketitle

\begin{abstract}
Static and dynamic equilibria in noisy binary choice games on graphs are considered. Equations defining static quantal response equilibria (QRE) for binary choice games on graphs with arbitrary topology and noise distribution are written. It is shown that in the special cases of complete graph and arbitrary noise distribution, and circular and star topology and logistic noise distribution the resulting equations can be cast in the form coinciding with that derived in the earlier literature. Explicit equations QRE for non-directed graphs in the annealed approximation  are derived. It is shown that the resulting effect on the phase transition is the same as found in the literature on phase transition in the Ising model on graphs in the same approximation.Evolutionary noisy binary choice game having the earlier described QRE as its stationary equilibria in the mean field approximation is constructed using the formalism of master equation.

\end{abstract}

\newpage

\section{Introduction}

Taking into account direct not market-mediated economic/social interactions, i.e. a direct dependence of agent's utilities on the (expected) actions of fellow agents, is believed to be of crucial importance for describing the observed heterogeneity of economic/social multi-agent systems. The corresponding literature that contains many practical applications of this idea is covered in the review papers \cite{blume2001interactions,durlauf2010social}. Particularly interesting consequences of such interactions are emergent nontrivial aggregate properties of these systems. 

At the same time the direct dependence of agent's utilities on the (expected) actions of other agents is, by default, a key ingredient of game theory. It is therefore natural to look at the consequences of economic/social interactions in terms of emergent aggregate properties of game-theoretic equilibria.

Of direct relevance to the present paper is the literature on effects of social/economic interactions on equilibria properties in noisy binary choice problems on graphs with utility of a choice containing private and social components where the former depends on idiosyncratic fixed characteristics and noise and the latter depends on (expected) choices made by the agent's neighbours. The resulting description depends on the assumptions on the form of this interaction, topology of a graph and specification of noise. Static equilibria and corresponding myopic dynamics for noisy binary choice problem on complete graph were considered, for linear-quadratic utility and arbitrary noise distribution, in  \cite{brock2001discrete,blume2003equilibrium}.  
The particular case of linear-quadratic utility and logistic noise was considered for several topologies, including complete graph, star and circular ones\footnote{The case of the circular topology was also considered in the influential early paper \cite{glaeser1996crime}.} in \cite{ioannides2006topologies}. The main result of this literature is in describing nontrivial emerging aggregate properties of equilibria which in some special cases resemble or coincide with description of phases and phase transitions in statistical physics. In particular, in the case of logistic noise and complete graph and circular topologies the resulting equations describing static equilibria and dynamical evolution coincide with those describing magnets in the mean field approximation. An inspiring discussion of parallels between noisy discrete decision problems in multiagent systems and statistical physics can be found in \cite{bouchaud2013crises}, see also  \cite{durlauf2018statistical}.

As has been already mentioned, a natural language for developing a consistent description of the effects of economic/social interactions of a group of agents is provided by game theory in which interdependence of agent's strategies is built in by construction through their utilities. The goal is then to describe the corresponding game-theoretic equilibria. The randomness present in agent's decisions makes it necessary to describe these equilibria in probabilistic terms. In the general case of interdependent random effects influencing agent's decisions these are correlated equilibria \cite{aumann1987correlated}. In the simplest case in which such random effects are assumed to be independent one looks for equilibria in mixed strategies. Let us note that in game-theoretic literature randomness in agent's utilities is considered as a possible fundamental mechanism underlying formation of equilibria in mixed strategies \footnote{An alternative explanation is through bounded  rationality, see e.g. \cite{chen1997boundedly}.} \cite{harsanyi1973games}. 

Of particular relevance to the present study is a specific implementation of the Harsanyi random utility mechanism \cite{harsanyi1973games} resulting in particular Bayes-Nash mixed strategies equilibria in games with incomplete  (due to randomness in agent's utilities) information  - the Quantal Response Equilibria (QRE) \cite{mckelvey1995quantal,mckelvey1996statistical,goeree2016quantal}. An important result of \cite{mckelvey1995quantal} was in deriving explicit equations for equilibrium probability measure that were subsequently used for describing many experimental game-theoretic results \cite{goeree2016quantal}.

The main goal of this paper is to describe QRE equations and the properties of the corresponding equilibria for games corresponding to the noisy binary choice model  of  \cite{brock2001discrete} for graphs with arbitrary topology and noise distributions in both static and dynamic settings \footnote{Let us note that although the relevance of QRE in the context of discrete choice models was already mentioned in the literature, see e.g. \cite{ioannides2006topologies,durlauf2010social}, the explicit interrelation was not discussed.}. We show that in the particular cases of complete, star and circular graph topology  the QRE - defining equations  are equivalent to the equilibrium-defining equations in \cite{brock2001discrete} and, for a particular choice of noise distributions, in \cite{ioannides2006topologies}. We also consider QRE for noisy binary choice games on random graphs in the annealed approximation and derive an analytical equation defining them that, for a particular choice of logistic noise distribution, coincides with the one known in statistical physics on graphs \cite{dorogovtsev2008critical}. For dynamic binary choice games we derive, using the formalism of master equation, evolution equations for local average choice and show that static QRE equilibria constitute stationary points of a particular myopic stochastic choice game considered in the mean field approximation. In the particular case of a complete graph the resulting equation coincides with the one derived in \cite{blume2003equilibrium}.

\section{Quantal response equilibria in static noisy binary choice games}\label{section:static_game}

\subsection{Noisy binary choice game}

Let us define a noisy binary choice game considered in the paper:
\begin{itemize}
\item the game is played by $N$ agents placed in the vertices of a graph ${\cal G}$ characterised by the adjacency matrix $g$ with $g_{ij}=1$ corresponding to an edge $j \to i$ and $g_{ij}=0$ otherwise;
\item Each agent solves a binary choice problem described by two alternative pure strategies parametrised by $s_i = \pm 1$, $i=1, \cdots, N$;
\item the expected utility for an agent $i$ from choosing the strategy $s_i$ is assumed to depend on his expectations with respect to strategies chosen by his first neighbours   $\{j \in \nu_i \}$ and an idiosyncratic random contribution: 
\begin{equation}\label{equation:static_choice}
\langle U_i (s_i) \rangle =   \left[ H_i + \sum_{j \in \nu_i} g_{ij} J_{ij}   \langle s_j \rangle_i  \right ] s_i  + \epsilon_{s_i},
\end{equation}
where $H_i$ is an idiosyncratic bias towards choosing $s_i = {\rm sign} (H_i)$, the social/economic interactions are parametrised by the matrix $J_{ij}$ describing the influence of the choice of strategy $s_j$ by the neighbour $j$. The random contributions $\{ \epsilon_{s_i} \}$ are assumed to be privately known independent (for $i \neq j$)) random variables with zero mean.  For each agent $i$ distributions $\phi^{(i)} (\epsilon_{s_i})$ for 
$\epsilon_{s_i}$ and  $\epsilon_{-s_i}$  are assumed to be {\it the same}\footnote{This assumption is necessary to be able to falsify predictions of the properties of quantal response equilibrium, see \cite{haile2008empirical}.}. In what follows we denote this distribution by $\phi^{(i)} (\epsilon_{s_i})$. The expectation values $ \langle s_j \rangle_i  $ are average choices of the agents $\{j \in \nu_i \}$ computed following agent's $i$ evaluation of the corresponding probabilities. The choice of linear-quadratic form of the utility \eqref{equation:static_choice} is standard and follows, in particular,  the one in \cite{brock2001discrete,blume2003equilibrium}. 
\end{itemize}

\subsection{Non-random case}

Let us first consider the "unperturbed" version of \eqref{equation:static_choice} with the noise terms switched off:
\begin{equation}\label{equation:static_choice_unpert}
U_i (s_i)  =   \left[ H_i + \sum_{j \neq i} g_{ij} J_{ij}   s_j  \right ] s_i.
\end{equation}
The Nash equilibria in pure strategies  $(s^*_1, \cdots,s^*_N)$ correspond to solutions the system of equations characterising the best response strategies of the agents
\begin{equation}\label{equation:NE_pure_strategies}
s^*_i  = \eta \left( H_i + \sum_{j \neq i} g_{ij} J_{ij}   s^*_j   \right),
\end{equation}
where $\eta(x) = {\rm sign}(x)$ is a step function. Let us stress that an existence of solutions of \eqref{equation:static_choice_unpert} is in general case not guaranteed and conditions of their existence are not known. As for the Nash equilibria in mixed strategies, to our knowledge  
they were not discussed. The interpretable solutions of the system  \eqref{equation:NE_pure_strategies} can be found only in some drastically simplified versions of the game under consideration. For example, for the case of complete graph, $H_i=0 \; \forall\;  i$ and $J_{ij} = J/N >0 \; \forall \; i,j$ the two Nash equilibria in pure strategies are $s^*_i = 1 \; \forall \; i$ or $s^*_i = -1 \; \forall \; i$.  In the particular case of symmetric matrices $g_{ij}$ and $J_{ij}$ the game is a potential one, see e.g. \cite{le2011games}, with the potential of the form
\begin{equation}
V \left( \{ \sigma \}\right) = \sum_i H_i \sigma_i + \frac{1}{2} \sum_{i,j} g_{ij} J_{ij}  s_i s_j 
\end{equation}
and, as proven in \cite{le2011games}, possesses therefore at least one Nash equilibrium in pure strategies. 

\subsection{Quantal response equilibrium}

Taking into account the noise terms in \eqref{equation:static_choice} requires characterising the corresponding equilibria in probabilistic terms. In the general case in which random utility components $\{ \epsilon_{s_i} \}$ are {\it dependent} such equilibria are characterised by multinomial probability distributions 
\begin{equation}\label{equation:static_distribution_function}
P^{\rm eq} \left( \{ s \} \right) = P^{\rm eq} \left( s_1, s_2, \; ... \; s_N \right)
\end{equation} 
denoting the probability of having an equilibrium characterised by the set of strategies $(s_1, s_2, \cdots s_N)$. Such equilibria can be classified as correlated  \cite{aumann1987correlated}. 

With the above-described standard assumption of independence of $\{ \epsilon_{s_i} \}$ for different agents used in the subsequent analysis one gets a factorised distribution 
\begin{equation}\label{equation:static_factorised_distribution_function}
P^{\rm eq} \left( \{ s \} \right)  = P^{\rm eq}_1 (s_1) P^{\rm eq}_2 (s_2) \; ... \; P^{\rm eq}_N (s_N)
\end{equation} 
corresponding to Nash equilibrium in mixed strategies parametrised by the corresponding noise-induced set of equilibrium probabilities $\{p^{\rm eq}_{s_i} \}$, where  $\{ p^{\rm eq}_{s_i}  \equiv P^{\rm eq}_i (s_i) \}$ and  $\{ p^{\rm eq}_{s_i=1}+ p^{\rm eq}_{s_i=-1}=1 \}$. %In what follows we will confine ourselves to analysing the particular version of mixed strategy equilibria - the quantal response equilibria (QRE) \cite{mckelvey1995quantal,mckelvey1996statistical,goeree2016quantal}.

The agent's choices are now governed by expected utility \eqref{equation:static_choice}. From \eqref{equation:static_choice} it follows that the probability $p_{s_i}$ of choosing the strategy $s_i$ by the agent $i$ is 
\begin{equation}\label{equation:static_probability}
p_{s_i} = F_<^{(i)} \left( \left[ 2 H_i + 2 \sum_{j } g_{ij} J_{ij}  \langle s_j \rangle_{(i)} \right] s_i \right),
\end{equation} 
where $F_<^{(i)} (z)$ is a distribution function for $\epsilon_{-s_i} - \epsilon_{s_i}$ 
\begin{equation}
 F_<^{(i)} = {\rm Prob} \left[ \epsilon_{-s_i} - \epsilon_{s_i} < z \right] = \int^z dz_1 f^{(i)} (z_1), 
\end{equation}
where, in turn, $f^{(i)} (z_1) = \int dz_2 \; \phi^{(i)} (z_2) \phi^{(i)} (z_2+z_1)$ is the differential distribution of $z_1=\epsilon_{-s_i} - \epsilon_{s_i}$ \footnote{It is easy to check that the probabilities \eqref{equation:static_probability} satisfy $p_{s_i} + p_{-s_i} = 1$.}.
In particular, for the probability $p_{s_i=1}$ of choosing $s_i=1$ one has
\begin{equation}
p_{s_i=1} \equiv p_{+i} = F_<^{(i)} \left( 2 H_i + 2 \sum_{j } g_{ij} J_{ij}  \langle s_j \rangle_{(i)} \right),
\end{equation} 

The quantal response equilibrium $(p^{\rm eq}_{+1}, \cdots, p^{\rm eq}_{+N})$ is then defined \cite{mckelvey1995quantal,mckelvey1996statistical,goeree2016quantal} as a consistent set of mixed strategies used by agents and their expectations concerning the mixed strategies used by their neighbouring agents, i.e. fullfilment for $\forall j,i$  of the equality
\begin{equation}
\langle s_j \rangle_{(i)} = 2p^{\rm eq}_{+j}-1
\end{equation}
which then leads to the following system of equations for the QRE probabilities $\{ p^{\rm eq}_{+i} \}$:\footnote{An expression of this sort containing some arbitrary linear combination of probabilities in the right-hand side is provided,  without reference to QRE, in \cite{durlauf2010social} .}
\begin{equation}\label{equation:QRE_prob}
p^{\rm eq}_{+i} = F_<^{(i)} \left( 2 H_i + 2 \sum_{j } g_{ij} J_{ij}  (2p^{\rm eq}_{+j}-1) \right), \;\;\;\ \forall i .
\end{equation}
It is convenient to rewrite \eqref{equation:QRE_prob} as a system of equations on the equilibrium expectations/local averages $m^{\rm eq}_i = 2p^{\rm eq}_{+i}-1$:
\begin{equation}\label{equation:QRE_av}
m^{\rm eq}_i = 2  F_<^{(i)} \left( 2 H_i + 2 \sum_{j } g_{ij} J_{ij}  m^{\rm eq}_j \right) -1
\end{equation}

\subsubsection{Simple topologies}

Let us consider the particular realisation of equations \eqref{equation:QRE_av} for the simplest characteristic topologies of complete, star-like and circular graphs. In this section we will assume for simplicity that for complete graph $J_{ij} = J/N$, for star-like and circular ones  $J_{ij} = J$ and $F_<^{(i)} (z)=F_< (z)$ for all $i,j$.

The appearance of the $1/N$ factor in $J_{ij} = J/N$ for the complete graph and the absense of analogous rescalings for star-like and circular topologies requires a special comment. In statistical physics such rescaling is necessary for ensuring additivity of free energy in the number of spins. In generic statistical physics models on nontrivial graphs no rescaling with respect to the number of nearest neighbours like that assumed in \cite{ioannides2006topologies,durlauf2010social} is imposed. In particular, the absence of such rescaling is crucial for obtaining a description of phase transitions for arbitrary topologies in the annealed approximation that coincides with the corresponding results obtained for phase transitions on graphs in statistical physics \cite{dorogovtsev2008critical}, see below.

\begin{enumerate}

\item {\bf Complete graph} \\ 
Let us first consider the case of the complete graph.  In this case equilibrium strategies of all agents are the same  $m^{\rm eq}_i=m^{\rm eq} \; \forall i$. The corresponding QRE is fully characterised, in the limit $N \to \infty$, by the corresponding simplification of the equation \eqref{equation:QRE_av}
\begin{equation}\label{equation:QRE_av_cg}
m^{\rm eq} = 2F_< \left( 2H + 2 J m^{\rm eq} \right) - 1,
\end{equation}
i.e. the Curie-Weiss equation obtained in  \cite{brock2001discrete,blume2003equilibrium}. It is thus established that the equilibria studied in  \cite{brock2001discrete,blume2003equilibrium} are the quantal response ones. 

Let us also mention that in the case $H=0$ the space of solutions of  \eqref{equation:QRE_av_cg}  undergoes  transformation ("phase transition") at 
\begin{equation}\label{equation:PT_point_cg}
4J f(0) = 1 
\end{equation}
such that at $4J f(0) < 1$ one has only the solution $m^{\rm eq}=0$ while at $4J f(0) > 1$ there appear additional solutions $\pm m^{\rm eq} \neq 0$. 
%As at a moment we do not have a description of the perturbed game as a potential one, the question of the stability/ranking of these equilibria can be analysed only in the framework of dynamical description, see below. 

\item {\bf Star-like graph} \\
In the star-like configuration one has two kinds of nodes: the central and periphery ones. Let us index the central node by $i=1$ and denote the average choice at $i=1$ by $m_1$ and for $i=2, \cdots, N$ by $m_{-1}$.  Then the equations \eqref{equation:QRE_av} take the form
\begin{eqnarray}\label{equation:QRE_av_sg}
m^{\rm eq}_1 & = & 2F_< \left( 2H + 2 J [N-1] m^{\rm eq}_{-1} \right) - 1 \nonumber \\
m^{\rm eq}_{-1} & = & 2F_< \left( 2H + 2 J m^{\rm eq}_1 \right) - 1
\end{eqnarray}
Up to normalisation in the interaction term and for the special case of logistic noise distribution equations \eqref{equation:QRE_av_sg} coincide with those derived in \cite{ioannides2006topologies} and, therefore, the corresponding equilibria are QRE ones.

\item {\bf Circular graph} \\
For the circular topology the equations \eqref{equation:QRE_av} take the form
\begin{equation}\label{equation:QRE_av_cig}
m^{\rm eq}_i= 2F_< \left( 2H + 2 J [m^{\rm eq}_{i-1} + m^{\rm eq}_{i+1} ] \right) - 1
\end{equation}
which, again, up to normalisation in the interaction term and for the special case of logistic noise distribution equations \eqref{equation:QRE_av_cig} coincide with those derived in \cite{ioannides2006topologies} and, therefore, the corresponding equilibria are QRE ones.
\end{enumerate}

\subsubsection{Annealed approximation}

Let us now consider the quantal response equilibrium described by  \eqref{equation:QRE_av} on undirected graphs with arbitrary topology in the so-called annealed approximation, see e.g. the review  \cite{dorogovtsev2008critical}. In tis approximation the matrix elements $g_{ij}$ are replaced by probabilities of formation of an edge between the nodes  $i,j$ with degrees $k_i,k_j$ in the configuration model for random graphs
\begin{equation}\label{equation:adjmat_confmod}
g_{ij} \simeq \frac{k_i k_j}{N \langle k \rangle}
\end{equation}
and, therefore, an incomplete unweighted graph is transformed into a complete weighted one\footnote{Note the reappearance in \eqref{equation:adjmat_confmod} of the $1/N$ factor.} with weights given by \eqref{equation:adjmat_confmod}. This is the only approximation allowing to get analytical description of critical phenomena on tree-like graphs with arbitrary degree distribution and therefore having applications to many problems in statistical physics of spin systems, epidemics, synchronization, etc. on graphs, see e.g. \cite{dorogovtsev2008critical}. The essence of this approximation is in assuming the same properties (e.g. average strategies) hold for all agents placed in the nodes of the same degree and, therefore, providing a compact account of the node' s heterogeneity. Let us note that by construction the annealed approximation \eqref{equation:adjmat_confmod} works better for high degree nodes and, due to the fact that degree distribution of a giant cluster, on which such critical phenomena are usually studied, is biased towards higher degrees, the accuracy of the annealed approximation is often quite reasonable.

The QRE-defining system of equations \eqref{equation:QRE_av}  can then be rewritten in the form
\begin{equation}
m^{\rm eq}_{k_i} = 2F_< \left( 2H + 2 J k_i \sum_k \frac{k \pi_k}{ \langle k \rangle} m^{\rm eq}_k \right) - 1,
\end{equation}
where $\{ \pi_k \}$ is the degree distribution of the graph ${\cal G}$ and
\begin{equation}\label{equation:QRE_wav_1}
m^{\rm eq}_k = \left. \langle \sigma_j \rangle \right \vert_{k_j = k}^{\rm eq} 
\end{equation}
denotes equilibrium averages for agents having $k$ direct neighbours. 
Equations \eqref{equation:QRE_wav_1}  can conveniently be rewritten as an equation on the weighted average
 \begin{equation}
m^{\rm eq}_w = \sum_k  \frac{k \pi_k}{\langle k \rangle} m^{\rm eq}_k,
\end{equation}
so that the corresponding generalized Curie-Weiss equation for $m_w$ takes the form
\begin{equation}\label{equation:QRE_wav_2}
m^{\rm eq}_w = \sum_k  \frac{k \pi_k}{ \langle k \rangle} 2F_< \left( 2H + 2 J k m^{\rm eq}_w \right) - 1
\end{equation}
Let us note that for  $H=0$ the transformation of solution space for  $m_w$ take place at
\begin{equation}\label{equation:PT_point_anap}
4J \langle k^2 \rangle f(0) = \langle k \rangle
\end{equation}
Equation \eqref{equation:PT_point_anap} generalizes the corresponding equation for the complete graph \eqref{equation:PT_point_cg} so that one has the following transformation of the condition on the phase transition point:
\begin{equation}
4J f(0) = 1  \,\,\, \Rightarrow \,\,\, 4J  f(0) = \frac{\langle k \rangle}{\langle k^2 \rangle}
\end{equation}
which is a well-known effect in critical phenomena on graphs in the annealed approximation \cite{dorogovtsev2008critical}.
 For the often considered case of logistic $F^<(z)$  the equation  \eqref{equation:PT_point_anap}  coincides with that obtained in the same approximation for the Ising model on graphs \cite{dorogovtsev2008critical}.

\section{Dynamics}\label{section:evolutionary_game}

Let us now consider an evolutionary version on the static noisy binary choice game on graphs described in the Section \ref{section:static_game}. The corresponding dynamics can generically be described as an evolution of the multinomial probability distribution 
\begin{equation}\label{equation:distribution_function}
P \left(\{ s  \} (t) \right) = P \left( s_1, s_2, \, ... \, s_N \vert  t \right)
\end{equation} 
describing the probability of observing a configuration of strategies  $\{ s  \} (t) \equiv (s_1, s_2, \, ... \, s_N) (t)$  at time $t$. The evolution is triggered by strategy revisions $s_i \to  - s_i $ and is fully specified by describing how do revision possibilities occur and the corresponding probabilities $\{ p_{-s_i} (t) \}$ for an agent $i$, once given a chance to reconsider the current choice at time $t$, to choose the strategy $s_i$.  

In describing the dynamical evolution of $P \left(\{ s  \} (t) \right)$ we will
assume that it is driven by the strategy-switching transitions $s_i(t) \to  -s_i
(t),$ so that in a given infinitesimal time interval one van have only one such
flip for some agent. The underlying picture is that at some small time interval
$(t,t+\Delta t)$ only one agent $i$ gets a possibility of strategy revision. The overall probability of strategy revision  $-s_i \to   s_i $ is thus 
%\begin{equation}\label{equation:strategy_flip}
%{\rm Prob} \left[ -s_i \to   s_i  \right]_{(t,t+\Delta t)}= \lambda_i (t) \, p_{s_i} \, P \left (s_1, \, ... \, -s_i, \, ... \, s_N \vert t \right )
%\end{equation}
\begin{equation}\label{equation:strategy_flip}
{\rm Prob} \left[ -s_i \to   s_i  \right]_{(t,t+\Delta t)}= \lambda_i (t,t+\Delta t) \, p_{s_i} 
\end{equation}
where $\lambda_i (t,t+\Delta t)$ is the probability of giving an agent $i$ possibility of
reconsidering the current strategy on time interval $(t,t+\Delta t)$. Moreover,
we assume that probability of strategy selection during the reconsidering is
independent from the current state. Assuming, analogously to \cite{blume2003equilibrium}, independent Poissonian generation of time points at which strategy reconsideration is possible we get, using \eqref{equation:strategy_flip}, the following evolution equation (master equation) (see e.g. \cite{van1992stochastic}):
\begin{eqnarray}\label{equation:master_equation}
\frac{d P \left( s_1, s_2, \, ... \, s_N \vert  t \right)}{d t}  &                                                                      = & \lambda \sum_i [ p_{s_i} P \left (s_1, \, ... \, -s_i, \, ... \, s_N \vert t\right ) - \nonumber \\
& &  \;\;  p_{-s_i} P \left (s_1, \cdots, s_i, \cdots s_N \vert t\right ) ] \label{eqev0}
\end{eqnarray}
where $\lambda$ is a constant describing the intensity of the Poisson process. From the master equation \eqref{equation:master_equation} there follow equations for the moments. In particular, the evolution equations for the local averages $m_i(t) = \langle s_i \rangle_{P( \{ s \} (t))}$ reads
\begin{equation}\label{equation:evolution_av}
\frac{d m_i (t)}{ dt }= - 2 \langle s_i p_{-s_i}  \rangle_{P( \{ s \} (t))}
\end{equation} 

The ultimate goal is in describing asymptotic regimes of \eqref{equation:master_equation} at $t \to \infty$, in particular the possibility of stationary asymptotic regimes such that
\begin{equation}\label{equation:evolution_aymptotic}
\left. \frac{d P \left( \{ s \} (t) \right)}{d t} \right \vert_{t \to \infty} \to 0
\end{equation}
so that the evolution ends up with forming a stationary equilibrium distribution
\begin{equation}
\left. P \left( \{ s \} (t) \right) \right \vert_{t \to \infty} \; \to \; P^{\rm eq} \left( \{ s \} \right)
\end{equation}
In the present study we restrict our analysis to the case of mixed strategy equilibria in which agent's decisions are taken independently which in terms of the probability distribution $P \left( \{ s \} (t) \right)$ enforces using the factorized mean field approximation :
\begin{equation}\label{equation:distribution_function_factorized}
P \left( \{ s \} (t) \right)  \;\;\; \Rightarrow \;\;\; P_{\rm mf} \left( \{ s \} (t) \right) = \prod_{i=1}^N p_{s_i}(t)
\end{equation}
and, therefore, stationarity of $P \left( \{ s \} (t) \right)$ does directly imply that of the choice probabilities $p_{s_i}(t)$ and, therefore, of the local averages 
$m_i(t) = \langle s_i \rangle_{p_{s_i} (t)}$:
\begin{equation}
\left. \frac{d P \left( \{ s \} (t) \right)}{d t} \right \vert_{t \to \infty} \to 0 \;\; \Longrightarrow \;\;
\left. \frac{d p_{s_i} (t) }{d t} \right \vert_{t \to \infty} \to 0 \;\; \Longrightarrow \;\;
\left. \frac{d m_i (t) }{d t} \right \vert_{t \to \infty} \to 0
\end{equation}

Mechanisms determining the flip rate can in principle include memory effects, i.e. dependence of $\{ p_{s_i} \} $ on past configurations of strategies, see e.g. \cite{bouchaud2013crises} and/or forward-looking behaviour, see e.g. \cite{blume1995evolutionary}. In this study we will use, following \cite{blume2003equilibrium} and the earlier literature, the simplest myopic assumption so that a decision on the strategy flip depends only only on the current configuration of strategies at neighbouring nodes. In analogy with \cite{blume2003equilibrium} the expression for the flip rate is then naturally chosen as a myopic (local time-dependent) version of \eqref{equation:static_probability}: 
\begin{equation}\label{equation:flip}
p_{-s_i} = F_<^{(i)} \left( - \left[ 2 H_i + 2 \sum_{j } g_{ij} J_{ij}   s_j  \right] s_i  \right).
\end{equation}
 
 The general evolution equations for  $m_i (t)$  \eqref{equation:evolution_av} can be further simplified by using the identity
\begin{equation}\label{equation:identity}
 p_{-s_i} = \frac{1}{2} \left[ 1-s_i (s_i p_{s_i} - s_i p_{-s_i}) \right]=\frac{1}{2} \left[ 1- s_i \langle s_i \rangle \right] 
\end{equation}
Then, using equations (\eqref{equation:evolution_av},\eqref{equation:flip},\eqref{equation:identity},\eqref{equation:distribution_function_factorized}), we get the following evolution equations for the local averages $m_i(t)$:
\begin{equation}\label{equation:moment_evolution}
\frac{d m_i (t) }{dt} = -\lambda \left \{ m_i (t) - \left[ 2  F_<^{(i)} \left( 2 H_i + 2 \sum_{j } g_{ij} J_{ij}  m_j (t) \right )-1 \right] \right \}.
\end{equation}
This equation is approximate, however its validity can be proven for some graphs
in the limit of large number of neighbours. The evolutionary dynamics equilibria correspond to the (stable) stationary points of \eqref{equation:moment_evolution}:
\begin{equation}\label{equation:stationary_points}
\left. \frac{d m_i (t) }{dt} \right \vert_{t \to \infty} = 0 \; \Longrightarrow \; 
m_i  = 2  F_<^{(i)} \left( 2 H_i + 2 \sum_{j } g_{ij} J_{ij}  m_j  \right)-1  
\end{equation}
which are exactly the QRE static equilibria of \eqref{equation:QRE_av}.
Therefore in the mean field approximation
\eqref{equation:distribution_function_factorized} the dynamic equilibria of the
above-described evolutionary game with the myopic choice probabilities
\eqref{equation:flip} are the QRE of the noisy binary choice game described in
the section \ref{section:static_game}. An important point here is that not all
stationary points of the dynamics \eqref{equation:moment_evolution} are stable.
The stability of dynamical equilibria can be straightforwardly analyzed by
expanding the right-hand side of \eqref{equation:moment_evolution} to the second
order in perturbations around the corresponding equilibrium values and checking
whether these are growing or diminishing in time. One can check that for
stability it is necessary that the real parts of all eigenvalues of stability matrix 
\begin{equation}
  S_{ij}=\delta_{ij}-4f^{(i)}\left( 2 H_i + 2 \sum_{j } g_{ij} J_{ij}  m_j  \right)g_{ij}J_{ij}
\end{equation}
are positive.

In the simplest case of a complete graph and assuming $J_{ij} = J/N$, $H_i=H$ and $F_<^{(i)} (z)=F_< (z)$ for all $i,j$ the evolution equation \eqref{equation:moment_evolution} takes the form
\begin{equation}
\frac{d m (t) }{dt} = - \lambda \left\{ m(t) - [2F_< (2H + 2 J m ) - 1] \right\}
\end{equation}
thus reproducing the answer obtained in the framework of population dynamics formalism in \cite{blume2003equilibrium}. In the nontrivial phase of the model with $H=0$ corresponding to $4 J f(0) >1$ it is easy to check that the solution $m=0$ is dynamically unstable, see analogous conclusion for the myopic dynamics for logistic $F_<(0)$ in \cite{brock2001discrete}.

For the star and circular topologies and logistic $F_<(z)$ the equations \eqref{equation:moment_evolution}  coincide, after appropriate coupling rescaling, with those derived in \cite{ioannides2006topologies}.

\section{Conclusions and outlook}

Let us summarize the main results obtained in the present study:
\begin{itemize}
\item Equations defining quantal response equilibria in noisy binary games on graphs with arbitrary topology and arbitrary noise distributions were derived both for probabilities and local averages. An equivalence to the known results in the case of complete graph, star and circular topologies was estableshed.
\item Equations defining quantal response equilibrium in the annealed approximation for the underlying graph topology were derived and corresponding phase transition points identified. For the logistic noise distribution the answer was shown to be equivalent to that obtained in the framework of Ising model on graphs studied in statistical physics.
\item Master equation for evolution of the multinomial probability distribution describing evolution of strategies configuration and the corresponding evolution equations for local averages were derived. 
\item It was shown that for a natural myopic dynamics the stationary points of dynamical evolution equations (some of them unstable) correspond to the static QRE equilibria.

\end{itemize}

The present study dealt only with the subset of important issues related to developing a concise game-theoretic understanding of noisy discrete choice games on graphs. An incomplete list of the problems to be analyzed includes studying correlated equilibria, multinomial choice for arbitrary noise, effects of violation of detailed balance, possibility of developing a potential game description, etc. 

\bibliography{Ising_QRE}

\end{document}